\documentclass[twocolumn,showpacs,preprintnumbers,prl,aps,amssymb,superscriptaddress]{revtex4}
\usepackage{graphicx}
\usepackage{dcolumn}
\usepackage{bm}

\begin{document}

\newcommand{\Co}{CeCoIn$_5$}
\newcommand{\La}{Ce$_{1-x}$La$_x$CoIn$_5$}
\newcommand{\ie}{{\it i.e.}}
\newcommand{\eg}{{\it e.g.}}
\newcommand{\etal}{{\it et al.}}


\title{Unpaired Electrons in the Heavy-Fermion Superconductor CeCoIn$_5$}

\author{M.~A.~Tanatar}
\email{tanatar@ims.ac.jp}
\altaffiliation[Permanent address:]{~Inst. Surface Chemistry, N.A.S. Ukraine, Kyiv, Ukraine.}
\affiliation{Department of Physics, University of Toronto, Toronto, Ontario, Canada}

\author{Johnpierre~Paglione}
\altaffiliation[Present address:]{~Department of Physics, University of California, San Diego, La Jolla, CA, USA.}
\affiliation{Department of Physics, University of Toronto, Toronto, Ontario, Canada}

\author{S.~Nakatsuji}
\affiliation{Department of Physics, Kyoto University, Kyoto 606-8502, Japan}

\author{D.~G.~Hawthorn}
\altaffiliation[Present address:]{~Department of Physics and Astronomy, University of British Columbia, Vancouver, British Columbia, Canada.}
\affiliation{Department of Physics, University of Toronto, Toronto, Ontario, Canada}

\author{E.~Boaknin}
\altaffiliation[Present address:]{~Department of Physics, Yale University, New Haven, CT, USA.}
\affiliation{Department of Physics, University of Toronto, Toronto, Ontario, Canada}

\author{R.~W.~Hill}
\altaffiliation[Present address:]{~Department of Physics, University of Waterloo, Waterloo, Canada.}
\affiliation{Department of Physics, University of Toronto, Toronto, Ontario, Canada}

\author{F.~Ronning}
\altaffiliation[Present address:]{~Los Alamos National Lab, MST-10 Division, Los Alamos, NM, USA.}
\affiliation{Department of Physics, University of Toronto, Toronto, Ontario, Canada}

\author{M.~Sutherland}
\altaffiliation[Present address:]{~Cavendish Laboratory, University of Cambridge, Cambridge, United Kingdom.}
\affiliation{Department of Physics, University of Toronto, Toronto, Ontario, Canada}

\author{Louis Taillefer}
\affiliation{Department of Physics, University of Toronto, Toronto, Ontario, Canada}
\affiliation{Regroupement Qu\'eb\'ecois sur les Mat\'eriaux de Pointe, D\'epartement de physique, Universit\'e de Sherbrooke, Sherbrooke, Canada}
\affiliation{Canadian Institute for Advanced Research, Toronto, Ontario, Canada}

\author{C.~Petrovic}
\affiliation{Department of Physics, Brookhaven National Laboratory, Upton, New York 11973, USA}

\author{P.~C.~Canfield}
\affiliation{Ames Laboratory and Department of Physics and Astronomy, Iowa State University, Ames, Iowa 50011, USA}

\author{Z.~Fisk}
\affiliation{Canadian Institute for Advanced Research, Toronto, Ontario, Canada}
\affiliation{Department of Physics, University of California, Davis, Davis, California 95616, USA}

\date{\today}


\begin{abstract}

Thermal conductivity and specific heat were measured in the superconducting state of the heavy fermion material Ce$_{1-x}$La$_x$CoIn$_5$. With increasing impurity concentration $x$, the suppression of $T_c$ is accompanied by the increase in residual electronic specific heat expected of a $d$-wave superconductor, but it occurs in parallel with a {\it decrease} in residual electronic thermal conductivity. This contrasting behavior reveals the presence of uncondensed electrons coexisting with nodal quasiparticles. An extreme multiband scenario is proposed, with a $d$-wave superconducting gap on the heavy-electron sheets of the Fermi surface and a negligible gap on the light, three-dimensional pockets.

\end{abstract}

\maketitle


Many heavy-fermion materials possess a novel form of superconductivity thought to originate from pairing by magnetic fluctuations, a mechanism most favorable in the vicinity of a continuous zero-temperature magnetic instability \cite{Mathur}, or quantum critical point (QCP). The heavy-fermion material CeCoIn$_5$, with quantum criticality \cite{Sidorov,Paglione,Movshovich-QCP} and the highest transition temperature $T_c$=2.3~K amongst heavy-fermion superconductors \cite{Petrovic}, is an excellent candidate to study the relationship between QCP and unconventional superconductivity.

Experimental studies of the superconducting state in CeCoIn$_5$ have revealed a plethora of unconventional properties consistent with the presence of line nodes in the superconducting gap \cite{NQR,Movshovich-unconv,Brown,PCS}. The presence of a four-fold anisotropy in the thermal conductivity  suggests an order parameter with $d_{x^2-y^2}$ symmetry \cite{Izawa-aniz}. Several observations, however, remain unexplained in this picture. Recent specific heat measurements \cite{specheat-aniz} show a four-fold anisotropy which points instead to a $d_{xy}$ nodal gap structure. The temperature dependence of the spin-relaxation rate saturates below 0.3~K \cite{NQR}, while that of the penetration depth does not follow the functional form expected for a $d$-wave gap \cite{Brown}. 

Doping with impurities provides a powerful route to the gap structure of \Co. Non-magnetic impurities are strong pair-breakers in unconventional superconductors \cite{Hirsh}, acting to both suppress $T_c$ and generate quasiparticles in the nodal regions of the superconducting gap. Because these quasiparticles have a residual density of states (DOS) in the $T\to 0$ limit, they can be characterized by the electronic specific heat $\gamma_{0S}\equiv C_{0S}/T$ and thermal conductivity $\kappa_{0S}/T$ in the $T \to 0$ limit. With increasing impurity concentration, the expectation for a gap with line nodes is: ~1) a roughly linearly increasing $\gamma_{0S}$ and 2) a constant ({\it i.e.} ``universal'') $\kappa_{0S}/T$ \cite{universal}. Observed in the {\it d}-wave superconductors YBa$_2$Cu$_3$O$_7$ \cite{Taillefer} and Bi$_2$Sr$_2$CaCu$_2$O$_x$ \cite{Behnia}, and the spin-triplet superconductor Sr$_2$RuO$_4$ \cite{Suzuki}, universal conductivity is considered one of the most reliable tests of nodal structure, and hence order parameter symmetry \cite{Millis}.

In this Letter, we report a remarkable observation: while the increase of $\gamma_{0S}$ with $x$ in \La\ is in agreement with expectations for impurity pair-breaking in a $d$-wave superconductor, the large residual linear term $\kappa_{0S}/T$ measured at $x=0$ is not. Instead, rather than remaining constant (universal), it is found to {\it decrease } upon doping, tracking the suppression of {\it normal state} conductivity. We show that this contrasting behavior can be understood if a sizable fraction of conduction electrons in \Co $~$fails to participate in the superconducting condensation, and we attribute it to a negligible superconducting gap on part of the Fermi surface.


Single crystals of Ce$_{1-x}$La$_x$CoIn$_5$ were grown by the self-flux method \cite{Petrovic,Satorudoping} for $0<x<0.15$. Samples for transport measurements were cut into rectangular parallelepipeds with typical dimensions $\sim 2 \times 0.2 \times 0.1$~mm$^3$. Four-wire contacts with $\sim 5$~m$\Omega$ resistance were made with indium solder. Four-probe measurements of the electrical resistivity $\rho$ and thermal conductivity $\kappa$ were performed down to 50~mK, the latter with a standard 1-heater, 2-thermometer technique. Specific heat measurements were performed down to 300~mK in a Quantum Design PPMS. 


\begin{figure}
\centering
\includegraphics[totalheight=2.7in]{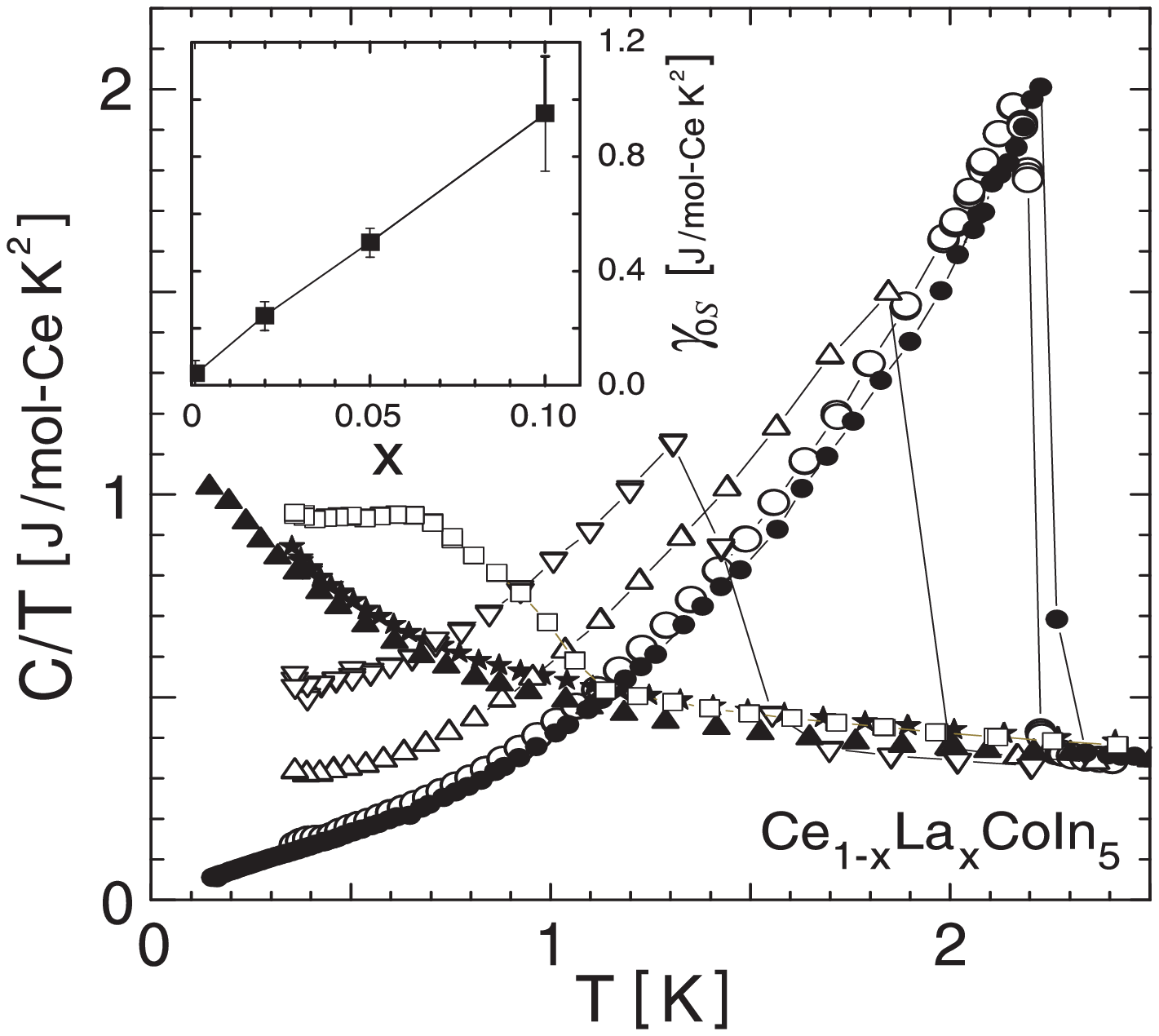}
\caption{\label{fig:doping} 
Temperature dependence of specific heat $C$ of Ce$_{1-x}$La$_x$CoIn$_5$ for $x$=0 (open circles), 0.02 (open up-triangles), 0.05 (open down-triangles), 0.1 (open squares), and 0.15 (solid stars) in zero magnetic field plotted as $C/T$ vs $T$. The data of Petrovic et al. \cite{Petrovic} for pure CeCoIn$_5$ in zero field (solid circles) and at 5 T (solid up-triangles) are shown for comparison.  Inset shows residual $\gamma_{0S}$ as a function of $x$.}
\end{figure}


The electronic specific heat $C$ of Ce$_{1-x}$La$_x$CoIn$_5$ in zero field is plotted as $C/T$ vs. $T$ in Fig.~\ref{fig:doping}. For the pure sample ($x=0$), the large jump of $C/T$ at $T_c$ is evidence for the participation of heavy-mass carriers in the superconducting condensation \cite{Petrovic}. Upon doping, the $T_c$, determined from the position of the $C/T$ peak, is gradually suppressed from 2.3~K at $x=0$ to 0.7-0.8~K at $x=0.10$ \cite{Satorudoping}. This is accompanied by a rapid increase in the normal state impurity scattering: $\rho_0$ measured at $H=10$~T (not shown) increases from $0.2~\mu\Omega$~cm at $x=0$ to $\sim 10~\mu\Omega$~cm at $x=0.10$. Such a decrease of $T_c$ implies a notable pairbreaking effect of nonmagnetic La impurities, which is characterized by a scattering rate $\Gamma \propto \rho_0 \propto x$ in the normal state \cite{Petrovic-impurities,NFP}. (The fact that $\rho_0$ varies linearly with $x$, as confirmed in the present study, shows that the Fermi surface is unchanged with increasing $x$, at least up to 15 \%.) For comparison, we show previously published data for $x=0$ \cite{Petrovic}, in zero field and in the field-induced normal state (5~T). The suppression of the superconducting state at $x=0$ by field (5~T) and by 15\%~La substitution yield identical $C/T$ curves. This further confirms that the normal-state density of states is unaffected by La doping, at least up to 15\%. As shown in the inset of Fig.~\ref{fig:doping}, $\gamma_{0S}$ increases linearly with $x$ \cite{determination}, as expected for non-magnetic impurities in unconventional superconductors \cite{Hotta,Maki}.

\begin{figure}
\centering
\includegraphics[totalheight=2.9in]{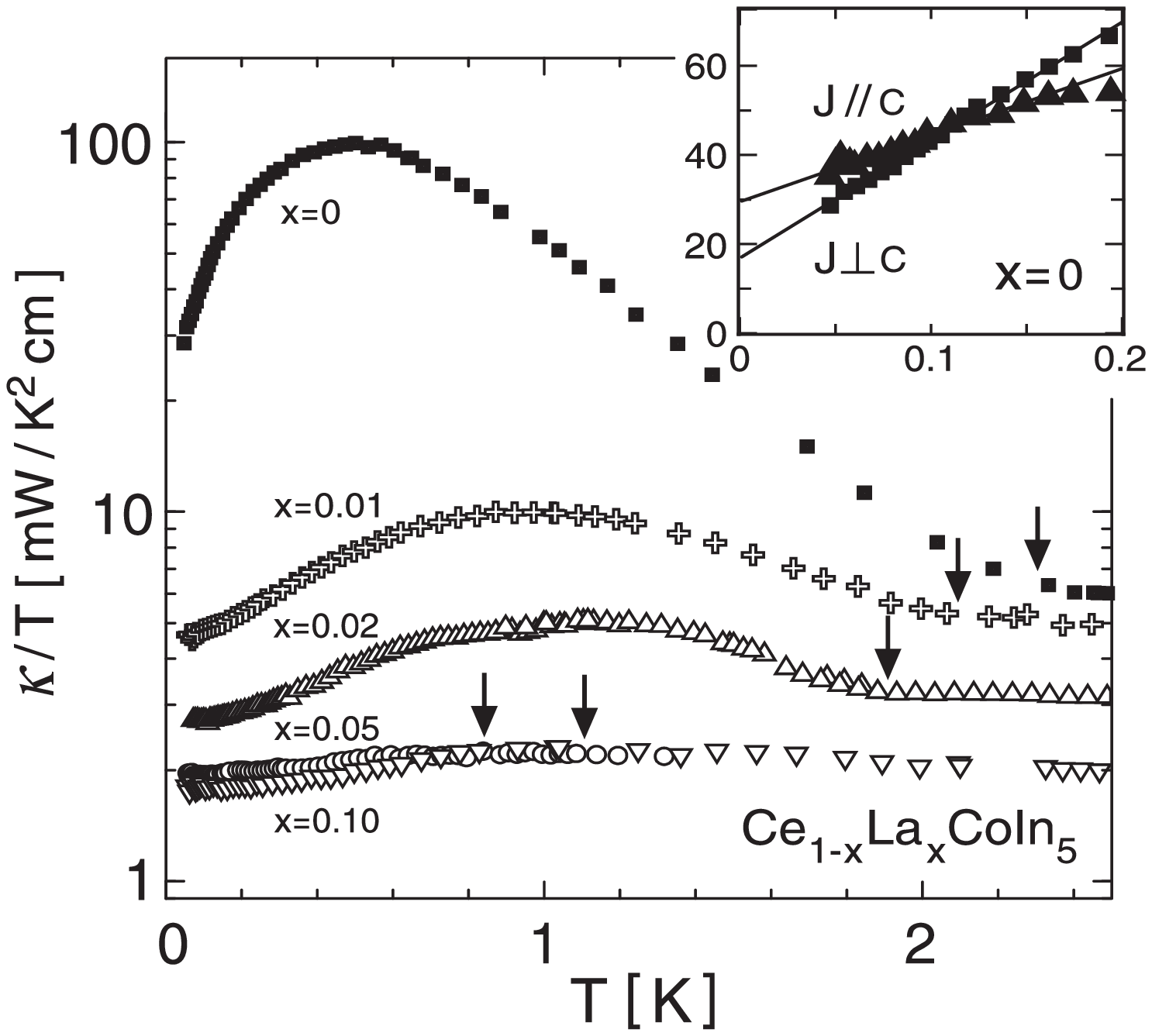}
\caption{\label{fig:kappa} Thermal conductivity $\kappa$ of Ce$_{1-x}$La$_x$CoIn$_5$ with $x$=0, 0.01, 0.02, 0.05 and 0.1 (top to bottom) in zero magnetic field, plotted as $\kappa/T$ vs. $T$ (note logarithmic scale). Arrows indicate $T_c$. Inset shows determination of residual linear term $\kappa_{0S}/T$ of pure ($x=0$) samples for in-plane (squares) and inter-plane (triangles) heat flow directions (linear scale).}
\end{figure}


In Fig.~\ref{fig:kappa} we present the thermal conductivity of Ce$_{1-x}$La$_x$CoIn$_5$ in the superconducting state for currents directed in the basal plane, plotted as $\kappa/T$ vs. $T$ on a semi-log scale. In our pure sample, $\kappa/T$ is roughly similar to previous reports \cite{Movshovich-unconv,Izawa-aniz}, increasing rapidly below $T_c$, reaching a maximum at $\sim 0.5$~K and then decreasing towards $T=0$. However, below $\sim 0.5$~K the behavior of $\kappa/T$ is notably different from that obtained in a previous study with $J \bot c$ \cite{Movshovich-unconv}, which found $\kappa \propto T^{3.4}$ for $T<100$~mK and determined $\kappa_{0S} /T$ to be less than 2~mW~K$^{-2}$cm$^{-1}$. In our experiment, $\kappa /T \propto T$ at low temperature and $\kappa_{0S} /T$ tends to a much larger residual value of $\sim 17$~mW~K$^{-2}$cm$^{-1}$, as shown in the inset of Fig.~\ref{fig:kappa}. Because it is known that the measurement of $\kappa$ can be inhibited by electron-phonon decoupling \cite{MikeSmith}, we have checked that our data indeed exhibits the intrinsic behavior by recovering the Wiedemann-Franz law in the normal state ($H>H_{c2}$) at $T\to 0$ \cite{contacts}. 

\begin{figure}
\centering
\includegraphics[totalheight=2.7in]{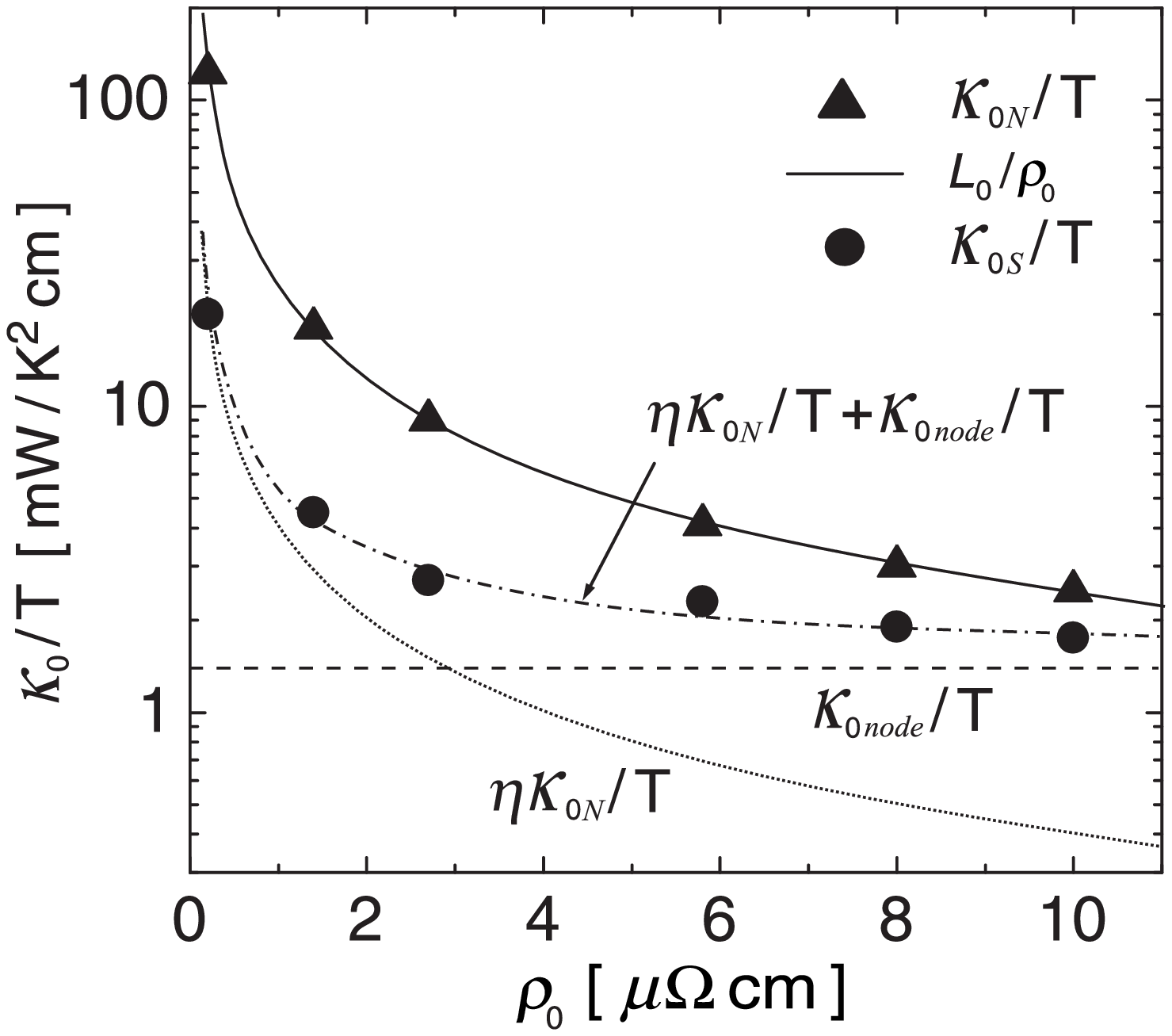}
\caption{\label{fig:universal} Residual ($T\to0$) electronic thermal conductivity in the normal ($\kappa_{0N}/T$, $H>H_{c2}$, triangles) and superconducting ($\kappa_{0S}/T$, $H$=0, circles) states of \La\ as a function of the normal state residual resistivity $\rho_0$ \cite{footnoterho}. The universal behaviour expected of a superconductor with line nodes is shown as a dashed line ($\kappa_{0node}/T$). The dashed-dotted line is a sum of uncondensed (dotted line) and nodal (dashed line) contributions, $\eta \kappa_{0N}/T + \kappa_{0node}/T$, with $\eta$=0.16.}
\end{figure}


Heat capacity and heat conduction are two diagnostic properties of a $d$-wave superconductor, and their essential inter-relation is captured by simple kinetic theory applied to the residual normal fluid of nodal quasiparticles: $\kappa_{0S}/T \propto \gamma_{0S} v_F^2 / \Gamma$, with $\Gamma \sim \rho_0 \sim x$, the impurity concentration. If the superconducting state has a DOS which varies linearly with energy, then one expects $\gamma_{0S} \propto x$, as is indeed observed here (see inset of Fig.~\ref{fig:doping}), and the two $x$ dependences (of $\gamma_{0S}$ and $\Gamma$) cancel out to leave $\kappa_{0S}/T$ universal, independent of $x$. A full theoretical treatment confirms this simple analysis and predicts universal heat conductivity \cite{universal}. In the normal state, the density of quasiparticle states $\gamma_{0N}$ remains constant and $\kappa_{0N}/T \sim \frac{1}{x}$. In Fig.~\ref{fig:universal}, we plot $\kappa_{0S}/T$ and $\kappa_{0N}/T$ as a function of $\rho_0(x)$ (determined for $H>H_{c2}$ in the $T\to 0$ limit) \cite{footnoterho}, and compare these to the behaviour expected of a superconductor with a line node in the gap: 1) universal limit in the superconducting state (constant $\kappa_{0S}/T$), and 2) Wiedemann-Franz law in the normal state  ($\kappa_{0N}/T$=$L_0/\rho_0$). The disagreement is dramatic. Rather than remaining roughly constant with increasing disorder, $\kappa_{0S}/T$ in fact decreases by a factor 5, much like the normal state conductivity. In Fig.~\ref{fig:universal} we show how a two-fluid scenario can capture this unusual behaviour: a band of electrons remains uncondensed below $T_c$, while the other is gapped with a line node.

We assume that some fraction $\eta$ of $\kappa_{0N}/T$ is preserved in the superconducting state, in addition to the contribution of nodal quasiparticles $\kappa_{0node}/T$, so that $\kappa/T = \eta \kappa_{0N}/T + \kappa_{0node}/T$. Assuming that $\kappa_{0node}/T$ obeys the universal limit and stays constant in the doping range studied \cite{contrib}, values of $\eta =  0.16$ and $\kappa_{0node}/T=1.4$~mW~cm$^{-1}$~K$^{-2}$ give a reasonable description of the data. The value we deduce for $\kappa_{0node}/T$ lies between published estimates of  the universal term for two-dimensional (1~mW~cm$^{-1}$~K$^{-2}$) and three-dimensional (1.9~mW~cm$^{-1}$~K$^{-2}$) superconducting states applied to CeCoIn$_5$  \cite{Movshovich-unconv}, assuming a $d$-wave gap \cite{universal}.  

The presence of uncondensed carriers suggests a multiband scenario, whereby at least one part of the Fermi surface does not participate in superconductivity. This is an extreme version of what has been established in a number of materials recently, where multiband superconductivity occurs either as a variation in the gap magnitude from Fermi surface sheet to Fermi surface sheet, as in MgB$_2$ \cite{MgB2}, or as a variation in nodal structure, as in Sr$_2$RuO$_4$ \cite{Deguchi_2004}. In the $s$-wave superconductor NbSe$_2$, for example, the gap maximum varies by a factor 3 or so \cite{Yokoya, Boaknin}. However, in no case so far has the gap been seen to actually vanish all the way to zero, as observed here in CeCoIn$_5$. The possibility of such an extreme case was predicted theoretically, however, by Agterberg et al. \cite{Agterberg1} in their model of orbital-dependent superconductivity, in which impurities scatter electrons predominantly within one band and not between bands \cite{Agterberg2}.

The value of $\eta$ we deduce suggests that the conductivity of the uncondensed carriers is 16\% of the total conductivity of all carriers (measured in the normal state). This is notably larger than their relative contribution to specific heat: in pure samples, $\gamma_{0S}~\approx$ 0.04~J~mole$^{-1}$K$^{-2}$ \cite{Movshovich-unconv}, or about 3\% of the normal state value in the $T\to 0$ limit (and it includes both nodal and uncondensed contributions). Therefore, the uncondensed carriers must either have a higher velocity than everage, or a lower scattering rate, or both. The ratio $v_F^2/\Gamma$ must be at least a factor 5 above average.


The calculated \cite{BandStructureCalc} and experimentally observed \cite{dHvA} Fermi surface of CeCoIn$_5$, allows for such large variation in carrier characteristics. The hole band 14 and electron band 15 form warped cylindrical surfaces with heavy carrier masses of $m^* \sim 30$-$100~m_0$, whereas the hole band 13 (and electron band 16 which is not observed in de Haas-van Alphen experiments) forms a three-dimensional surface with order-of-magnitude lighter carrier masses of $m^* \sim 4.3$-$12 ~m_0$. With a light mass, the contribution of the latter bands to conductivity can be much more important than their contribution to the DOS. 

Consequently, our proposal is that superconductivity in CeCoIn$_5$ originates on the cylindrical sheets of the Fermi surface, with negligible transfer of the order parameter onto the three-dimensional pockets. This interpretation is supported by two further experimental results. First, the {\it anisotropy} of thermal transport. As shown in the inset of Fig.~\ref{fig:kappa}, an anomalously large residual linear term is seen not only for a current in-plane, but also out-of-plane (along the c-axis), as you would expect from an {\it uncondensed} 3D Fermi surface. (In fact, $\kappa_{0S}/T$ is even larger for $J \parallel c$, namely 30~mW~cm$^{-1}$~K$^{-2}$.) Secondly, de Haas-van Alphen studies show that the $\epsilon$ orbit over the band 13 sheet exhibits negligible damping in the superconducting state, persisting down to $\sim 0.2 H_{c2}$ \cite{dHvA}, in contrast with strong damping observed for the $\alpha$ and $\beta$ orbits over band 14 and 15 sheets. In other words, holes in band 13 (and possibly electrons in band 16) do not feel the superconductivity. 

The presence of a group of light, uncondensed quasiparticles in the superconducting state can explain several experimental anomalies, such as the saturation of penetration depth at low temperatures \cite{Brown}. Orbital magnetoresistance of uncondensed carriers can explain the unexpectedly large difference in thermal conductivity \cite{Capan} between longitudinal and transverse field orientations in the plane with respect to the heat flow. A scenario of uncondensed electrons may also help resolve the apparent contradiction in the angular field dependence of thermal conductivity \cite{Izawa-aniz} and specific heat \cite{specheat-aniz}. Further implications of such a scenario for the nature of superconductivity in this material (and perhaps others) should be explored, e.g. the possibility of pure off-diagonal coupling \cite{Nichol}. Besides multiband superconductivity, other scenarios for uncondensed electrons should be considered, such as ``interior gap superfluidity,'' where a material is predicted to be simultaneously superconducting and metallic \cite{Wilczek}.


In conclusion, the qualitative contrast between low temperature specific heat and thermal conductivity in response to impurity doping in CeCoIn$_5$ has revealed the presence of a group of conduction electrons which do not participate in the superconducting condensation. This effect is naturally explained by assuming that one of the Fermi surfaces of CeCoIn$_5$ possesses a negligible superconducting gap. This is the most extreme case of multiband superconductivity encountered so far. Such an instance of complete decoupling between two groups of electrons in a single material will not only make all superconducting properties anomalous, it will also have profound implications for other cooperative phenomena, such as magnetic ordering and quantum critical behaviour. It provides a fascinating new window on electron correlations in metals. 


This work was supported by the Canadian Institute for Advanced Research and a Canada Research Chair (L. T.), and funded by NSERC. The authors are grateful to T. Hotta and H. Maehira for sharing the results of their band structure calculations, and to Y.~B.~Kim, I.~Mazin, K.~Behnia, Y.~Matsuda and M.~Smith for useful discussions. The work at Brookhaven was supported by the Division of Materials Sciences, Office of Basic Energy Sciences, U.S. Department of Energy under contract No. DE-AC02-98CH10886. The work at Kyoto university was supported by Grant-in-Aid for Scientific Research from JSPS. 



\begin{references}
\bibitem{contacts} The high conductivity of \Co\ places very strict requirements on the quality of thermal measurement contacts \cite{MikeSmith}. In samples with relatively high contact resistance, made following the procedure used in Ref.~\onlinecite{Movshovich-unconv}, we also observe a steep $\sim T^3$ dependence of $\kappa(T)$ and a small or negligible residual term $\kappa_{0S}/T$ in zero field. With such contacts, however, the normal-state ($H>H_{c2}$) conductivity, measured on the same samples, displays a large violation of the Wiedemann-Franz law, while this law is obeyed in low contact resistance samples.
\bibitem{footnoterho} We use $\rho_0$ as the variable instead of $x$ itself as this removes two sources of uncertainty, coming respectively from estimates of geometric factors (accurate to 10\% or so) and estimates of $x$ values. As mentioned earlier, $\rho_0 \propto x$ within error bars.
\end{references}
\end{document}